
\overfullrule=0pt
\parindent=24pt
\baselineskip=18pt
\magnification=1200


\def\ni{\noindent}
\def\op#1#2{{\displaystyle#1\above0pt\displaystyle#2}}
\def\rl{\rightline}
\def\cl{\centerline}

\def\gev{{\rm GeV}}

\def\e6{{\rm E}_6}
\def\su#1{{\rm SU(#1)}}

\def\NPB#1#2#3{Nucl. Phys. {B#1} (19#2) #3}
\def\PLB#1#2#3{Phys. Lett. {#1B} (19#2) #3}
\def\PRD#1#2#3{Phys. Rev. {D#1} (19#2) #3}
\def\PRL#1#2#3{Phys. Rev. Lett. {#1}, (19#2) #3}

\def\MPLA#1#2#3{Mod. Phys. Lett. {A#1} (19#2) #3}
\def\IJMPA#1#2#3{Int. J. Mod. Phys. {A#1} (19#2) #3}


{\nopagenumbers
\rl{CTP-TAMU-32/91}
\rl{May, 1991}
\vskip .5in
\cl{\bf Very Large Intermediate Breaking Scale}
\cl{\bf In The Gepner Three Generation Model}
\medskip
\cl{Jizhi WU\quad and\quad Richard ARNOWITT}
\centerline{\it Center for Theoretical Physics, Department of
Physics}
\cl{\it Texas A\&M University, College Station, TX77843, USA}
\vskip .5truein
\centerline{\bf Abstract}
\medskip
    A detailed study of the intermediate symmetry breaking scale, via the
renormalization group equations, for
a three generation heterotic string model arising
from the N=2 superconformal construction is reported. The numerical study
shows that the model admits a very large intermediate
breaking scale $\op{>}{\sim}1.0\times10^{16}$ GeV.
The role of the gauge singlets in this
model is studied, and it is found that these fields play a crucial role in
determining the directions and the scale of the intermediate symmetry breaking.
The importance of the mixing in generation space is also studied.
The generation mixing terms are found to have special effects in the
intermediate symmetry breaking. Remarkably these terms can produce
some {\it new} Yukawa couplings (not present at the Planck scale) through
loops. These couplings are in general very small compared to the ones
with non-vanishing tree level values and thus offer a {\it new}
mechanism to solve the lepton/quark mass hierarchy problem.
\medskip
\vfill
\eject}


\pageno=1
\cl{\bf 1.~~Introduction}
    Since the work of Green and Schwarz [1], the superstring theory has become
the best, if not the unique, candidate for the theory of the fundamental
interactions. The heterotic string [2] compactified on the Calabi-Yau manifolds
[3] emerged as the first phenomenologically viable theory in which the
internal degrees of freedom are given geometrical meaning. Here the
ten dimensional manifold $M_{10}$ is compactified to $M_4\times K_6$, where
$M_4$ is the Minkowski spacetime, and $K_6$ is a three
dimensional K\"ahler manifold with vanishing first Chern class (the
Calabi-Yau manifold). This programme provides us with an effective four
dimensional field theory of $N=1$
supergravity coupled to Yang-Mills theory with an $E_6\times E_8$ gauge
group [4].
The matter fields are in the fundamental 27 representations of $E_6$.

    On the other hand, superstring theories may be viewed as theories
on the two dimensional world sheet manifold whose properties are
governed by two dimensional superconformal field theories.
Here one does not need to adopt the geometrical point of view.
{}From the four dimensional Minkowski spacetime point of view, one can equally
well think of the theory as being constructed on $M_4\times S$, where
$S$ is taken to be the tensorial products of
discrete series of $N=2$ minimal superconformal field theories.
According to Gepner [5], one can take as many copies of $N=2$ models
as one likes,
subject to the condition that the total central charge of the tensorial
products adds up to 9, to
fulfill the anomaly cancellation requirement for the heterotic string.

    The Gepner construction of four
dimensional string theories is closely related to the Calabi-Yau models [6].
In particular, there is strong evidence that a superconformal construction
corresponds to a mirror pair of Calabi-Yau models with the betti numbers
$b_{11}$
and $b_{12}$ of the internal manifolds interchanged, {\it i.e.},
the family in one model is
actually the mirror family in the other and {\it vis versa} [7].
Although most work done along this line is in so-called weighted complex
projective manifolds of dimension four ($WCP^4$), and the
Tian-Yau manifold [3], the most studied three generation Calabi-Yau model [8],
has yet to be studied in superconformal field theory terms.
There does appear, however, a
three generation model which admits constructions both in terms of Calabi-Yau
manifold [9] and in terms of Gepner's language [10].
Hence, it is of interest
to study this model from both viewpoints.

    What makes Calabi-Yau models distinguished from the Standard Model
and Grand Unified Theories is that not only the
number of chiral fermion generations but also the Yukawa couplings among them
are determined by the topology of the internal Calabi-Yau manifold, though
subject to some still unknown normalizations of the fields [11]. The Yukawa
couplings among the 27's are not renormalized by any $\sigma$-model loop
corrections (hence are exact) due to a number of strong non-renormalization
theorems [12]. The calculability of Yukawa couplings greatly improves the
predictive
power of the Calabi-Yau models compared to the Standard Model or
the Grand Unified Theories in which one inserts
Yukawa couplings by hand to produce the masses appropriate for leptons
and quarks.
 Thus, even though there are still great
problems in determining the true vacuum of the heterotic string, the
low-energy spectrum and couplings are fixed once one chooses one's favourite
model.

    What makes Gepner's construction distinguished from Calabi-Yau models is
that not only the Yukawa couplings among the 27's of $E_6$,
but also those among the ${\overline{27}}$'s of $E_6$
(corresponding to (1,1)-forms of Calabi-Yau models) are calculable [13].
Moreover, all the nonrenormalizable interactions as well as the $E_6$ singlet
couplings
and the Yukawa couplings among them and chiral fields,
can be calculated, at least in principle, thanks to the intrinsic connection
between the $N=2$ minimal
models [14] and the parafermion models [15], and in turn, the SU(2) WZW
models [16] which has been studied in great detail [17].
The $E_6$ singlet couplings are less tractable in Calabi-Yau models because the
$E_6$ singlets belong to $H^1(End~T)$, the group of deformations of moduli, and
its dimension differs depending on the manifold. $E_6$ singlets play a very
important role in extracting low energy phenomenology [18]. As we shall see
later, they also play a unique role
in mediating the intermediate breaking of gauge symmetry in
the Gepner/Schimmrigk model which we will study in this paper.

    It should be emphasized that both these constructions or any other
approaches to four dimensional string theories still share a big drawback
with respect to supergravity theories in that the mechanism leading to the
breaking of supersymmetry is not yet understood.
This drawback strongly restricts the predictive power of the model.

This paper is organized as follows. In section 2, we briefly discuss the
construction of the Gepner three generation model [10] and
its Calabi-Yau counter
construction as given by Schimmrigk [9], mainly to set up notations and
conventions. Section 3 is devoted to the discussion of the renormalization
group equations of the model and the intermediate breaking scale.
Section 4 contains conclusions and possible future work that can be done.
A note about our notation: in this paper,
we shall not distinguish between the spinor and
the scalar components in the same matter multiplet, e.g., $L_8$ denotes both
spinor and scalar components of the eighth lepton generation.

\cl{\bf 2. Construction of the Model}
\noindent{\it Construction of the model}

    In constructing the three generation model, one starts by taking a
tensorial product of one level 1 and three level 16 models from the discrete
series of the $N=2$ superconformal theories [10]. Each field in the
$1\times(E16)^3$
model (where E means that we take $E_{7}$ invariant which exists for level 16
when constructing modular invariant partition function) consists of a product
of four `` atomic '' fields which are primary fields of the $N=2$ discrete
series. These are labelled by
$${\pmatrix{l&q&s\cr{\bar l}&{\bar q}&{\bar s}\cr}}_k,\eqno(1)$$
where $k$ is the level, $l~({\bar l})$, the ``isospin'', $q~({\bar q})$, the
``magnetic'' quantum number of the underlying left (right) SU(2) current
algebra, and $s~({\bar s})$ denotes the sector: $s~({\bar s})=0$, NS sector;
$s~({\bar s})=\pm1$, Ramond sector, and $s~({\bar s})=\pm2$ are the fields
obtained by acting on NS or R states by the supercharges, $G^{\pm}_{1/2}$,
of the $N=2$ superconformal algebra.

    Each primary field in Eq. (1) at level $k$ has a central charge
$c_k=3k/(k+2)$.
(Thus the $1\times(E16)^3$ model has a central charge 9, cancelling the
anomaly.) The conformal dimension and $U(1)$ charge are given by
$$h={l(l+2)-{\mathaccent "7E q}^2\over 4(k+2)}+{1\over8}s^2,\eqno(2)$$
$$Q=-{{\mathaccent "7E q}\over k+2}+{1\over2}s\eqno(3)$$
where ${\mathaccent "7E q}=q+s$.

    The modular invariant partition function can be constructed, by using the
$E_7$ invariant available at level 16, in a way that is consistent with
$N=1$ spacetime supersymmetry. One keeps only those composite states in
tensorial products with {\it odd integer} $U(1)$ charge (the generalized
GSO projection). The resultant model can be mapped onto a heterotic one.

    The model so constructed has a very large discrete symmetry. This can be
seen from the connection between the $N=2$ discrete series at level $k$ and the
$Z_{k}$ parafermion model --- they differ from each other by a trivial scalar.
This model has a $Z_{k+2}$ symmetry, which is just $Z_{18}$
for $k=16$ with a $Z_2$ factor being trivial. Hence, each copy of $E16$ field
in
the composite fields has a $Z_9$ symmetry.

    The resultant model has 35 generations (27's of $E_6$), 8 mirror
generations (${\overline{27}}$'s of $E_6$) and 197 scalars (singlets of $E_6$),
with
a discrete symmetry group
$${\mathaccent "7E G}=S_3\times G/Z'_9,\eqno(4)$$
where $Z'_9$ is the $Z_9$ subgroup of
$$G=Z_3\times {Z_9}^3,\eqno(5)$$
generated by the group element of $G$ carrying the $G$-charge (1,1,1,1).
$S_3$ is the permutation symmetry of three $E16$ factors. This is a
$35-8=27$ generation model. One can obtain a three generation model by
`` twisting '' the model by a $Z_3\times Z'_3$ subgroup of
${\mathaccent "7E G}$, with $Z_3$ being the cyclic permutation subgroup
of $S_3$ and $Z'_3$ generated by the $G$-charge (0,3,6,0).

    The above construction is related to a Calabi-Yau model due to
Schimmrigk [9]. The Calabi-Yau manifold is defined by two polynomials in the
ambient space $CP^3\times CP^2$ :
$$\sum_{i=0}^3 z_i^3=0;\qquad\qquad\sum_{i=1}^3z_ix_i^3=0\eqno(6)$$
where $z_i$, $i=(0,1,2,3)$, are the coordinates of $CP^3$, and $x_i$,
$i=(1,2,3)$, are those of $CP^2$. This manifold has the
Euler characteristic $\chi=-54$ and hence possesses 27 generations.
As shown by Gepner, this model has exactly the same generation structure and
discrete symmetry, Eq. (4), as the $1\times {E16}^3$ model constructed above.
One can obtain a three generation model by moding out two discrete
groups ${\mathaccent "7E Z}_3$ and ${\mathaccent "7E Z}'_3$ of order three,
$$\eqalign{{\mathaccent "7E Z}_3:\qquad g:\quad(z_0,z_1,z_2,z_3;x_1,x_2,x_3)
&\longrightarrow (z_0,z_2,z_3,z_1;x_2,x_3,x_1);\cr
          {\mathaccent "7E Z}'_3:\qquad h:\quad(z_0,z_1,z_2,z_3;x_1,x_2,x_3)
&\longrightarrow
(z_0,z_1,z_2,z_3;x_1,\alpha x_2,\alpha^2x_3),\cr}\eqno(7)$$
where $\alpha^3=1$, $\alpha\ne1$. Notice that ${\mathaccent "7E Z}_3$ is
freely acting, but ${\mathaccent "7E Z}'_3$ leaves three tori
$$\eqalign{(z_0^3+z_2^3+z_3^3&=0)\times(1,0,0);\cr
          (z_0^3+z_1^3+z_3^3&=0)\times(0,1,0);\cr
          (z_0^3+z_1^3+z_2^3&=0)\times(0,0,1)\cr}\eqno(8)$$
invariant. A smooth Calabi-Yau
manifold can be obtained by blowing up the invariant tori which will
necessarily introduce new massless modes whose Yukawa couplings
are not easy to calculate.

    The blowing-up procedure is analogous to the twisting in superconformal
construction. But the twisting is nothing more than assigning new boundary
conditions on massless modes and hence does not create any essential
difficulties for the calculation of correlation functions. This is another
advantage of Gepner's construction over Schimmrigk's.

\noindent{\it Massless modes and couplings}

The gauge group $E_6$ can be broken down to $[{\rm SU(3)}]^3$ {via}
flux-breaking.
As shown in [19], of two possible embeddings of $Z_3\times Z'_3$
into $E_6$, depending on whether $Z_3$ or $Z'_3$ correspond to
a trivial embedding.
The one corresponding to the trivial embedding of $Z_3$ introduces large
intrinsic CP violations [19] which we will not consider in this paper.
Thus the trivial embedding of $Z'_3$ generates 9 lepton,
6 mirror lepton, 3 quark, 3 anti-quark and no mirror quark or mirror
anti-quark generations [20, 19, 21], which, in terms of
${\rm SU(3)}_C\times{\rm SU(3)}_L\times{\rm SU(3)}_R$ quantum numbers,
can be represented by:
$$9L(1,3,{\bar 3})\oplus6{\bar L}(1,{\bar 3},3)\oplus3Q(3,{\bar 3},1)
  \oplus3Q^c({\bar 3},1,3)\eqno(9)$$
where $L\oplus Q\oplus Q^c$ furnish the 27's, and
${\bar L}\oplus{\bar Q}\oplus{\bar Q}^c$, the ${\overline{27}}$'s, of $E_6$.
In addition to the chiral generations, there are 61 gauge singlets, ${\phi}_i$,
where $i=1,...,35$, correspond to the 35 moduli fields, and $i=36,...,61$, to
the remaining singlets.
In terms of the SM quantum numbers, the nonets in Eq. (9) can be denoted by:
$$ (L_r^l)_i={\pmatrix{h^0&h^{+}&e^c\cr
                       h'^{-}&h'^0&\nu^c\cr
                       e&\nu&N\cr}}_i,\qquad i=1,...,9; \eqno(10)$$
$$ ({\bar L}_l^r)_{\bar i}={\pmatrix{
                       {\bar h}^0&{\bar h}'^{+}&{\bar e}\cr
                       {\bar h}^{-}&{\bar h}'^0&{\bar \nu}\cr
                       {\bar e}^c&{\bar \nu}^c&{\bar N}\cr}}_{\bar i},
   \qquad {\bar i}=1,...,6;\eqno(11)$$
$$ (Q_l^a)_i={\pmatrix{u^a\cr d^a\cr D^a\cr}}_i,
   \qquad i=1,2,3;\eqno(12)$$
$$ ({Q^c}_a^r)_i={\pmatrix{{u^c}_a&{d^c}_a&{D^c}_a\cr}}_i,
   \qquad i=1,2,3 \eqno(13)$$
where $a=1,2,3$, is the ${\rm SU(3)}_C$ index, $l=(e,\nu)$,
$q^a=\pmatrix{u^a\cr d^a\cr}$
$H=(h^0,h^{+})$, and $H'=(h'^{-},h'^0)$ are the lepton, quark and Higgs
doublets, $D^a$, ${D^c}_a$ are the color Higgs triplets of the SU(5)
supersymmetric grand unified theory, $N$ and $\nu^c$ are
SU(5) singlets while $N$ is also an SO(10) singlet. Finally, the fields
with a bar are the corresponding mirror fields.

    The Yukawa couplings of this model have been calculated by many
groups [22, 23, 24]. Here and in what follows, we use the results of [22]
which provides the most exhaustive list of Yukawa couplings as well as
possible non-zero nonrenormalizable interactions.
Our notations for gauge singlets also follow that of [22].
The most general $[{\rm SU(3)}]^3$ invariant cubic superpotential depends on
the
following Yukawa couplings:
$$\eqalign
     {{\bf W}_3
         =&{\lambda^1}_{ijk}det(Q_iQ_jQ_k)
         +{\lambda^2}_{ijk}det({Q^c}_i{Q^c}_j{Q^c}_k)\cr
         &+{\lambda^3}_{ijk}det(L_iL_jL_k)
         +{\lambda^4}_{ijk}Tr(Q_iL_j{Q^c}_k)\cr
         &+{\bar\lambda}_{{\bar i}{\bar j}{\bar k}}
           det({\bar L}_{\bar i}{\bar L}_{\bar j}{\bar L}_{\bar k})
         +{\eta}_{ij{\bar k}}\phi_iTr(L_j{\bar L}_{\bar k})
         +{\rho}_{ijk}(\phi_i\phi_j\phi_k),\cr}\eqno(14)$$
where in terms of the Standard Model particles in Eq. (10) -- (13):
$$\eqalign{det(QQQ)&=\epsilon_{aa'a''}d^au^{a'}D^{a''};\cr
     det(Q^cQ^cQ^c)&=\epsilon^{aa'a''}{d^c}_a{u^c}_{a'}{D^c}_{a''};\cr
           det(LLL)&=H^\lambda H'_\lambda N+H^\lambda\nu^c l_\lambda
                    +H'_\lambda e^c l^\lambda;\cr
          Tr(QLQ^c)&=D^aN{D^c}_a+D^ae^c{u^c}_a+D^a\nu^c{d^c}_a\cr
                   &+q^{a\lambda}l_\lambda{D^c}_a
                    +{q^a}_\lambda H^\lambda{u^c}_a
                    +q^{a\lambda}{H'}_\lambda{d^c}_a,\cr}\eqno(15)$$
here $\lambda$ is the ${\rm SU(2)}_l$ index, $a$, the ${\rm SU(3)}_c$
index and $i$, $j$, $k$, the generation index.
It turns out that there are 23 $(27)^3$, 3 $({\overline{27}})^3$,
25 $[1(27{\overline{27}})]$ and 110 $(1)^3$ non-vanishing Yukawa couplings, and
the structure of $(1)^3$couplings reveals that the moduli fields do not
couple among themselves [22]. Yukawa couplings of type $(27)^3$
and $({\overline{27}})^3$,
together with those of type $[1(27{\overline{27}})]$ which are relevant
for our discussion in the following sections are listed in Table 1. Notice
that, in Table 1, we also listed the values of these couplings at one loop
level, the results from running the renormalization group equations
to be discussed in Section 3.
The potentially non-vanishing non-renormalizable interactions which are not
forbidden by selection rules were also studied in Ref.~[23].

\centerline{\bf 3.~~Renormalization Group Equations and Large Intermediate
Breaking Scale}

    If we are to make any predictions for low energy phenomena at the
electroweak scale $M_{ew}\approx 100$ GeV,
the gauge group ${\rm SU(3)}_C\times{\rm SU(3)}_L\times{\rm SU(3)}_R$
must be broken to the Standard Model gauge group
${\rm SU(3)}_C\times{\rm SU(2)}_L\times{\rm U(1)}_Y$.
Moreover, supersymmetry must also be broken because of apparent asymmetry
between bosons and fermions at the electroweak scale. The origin of the
supersymmetry
breaking remains a most challenging problem in elementary particle physics
despite many efforts devoted to the issue. We do not attempt to make any
conjecture about the solution of this problem in this paper. Thus we shall
simply assume some mechanism exists to softly break supersymmetry at a
very high scale close to that of compactification which is not much
less than Plank scale, {\it i.e.}, we assume that
$M_{SUSY}\sim M_{C}\sim2.4\times10^{18}$ GeV.
Then the soft breaking of supersymmetry will introduce the standard soft
breaking terms:
\item{}{\it 1.} A common scalar mass for all spin zero modes;
\item{}{\it 2.} A common mass for all gauginos;
\item{}{\it 3.} Trilinear scalar couplings for all the Yukawa couplings.

    To break the gauge symmetry, we adopt {\it intermediate breaking
mechanism} by taking negative mass squared as the signal of the Higgs-like
mechanism [25]. Thus, we want to run the renormalization group equations for
({\it 1}) Gauge Couplings; ({\it 2}) Yukawa Couplings; ({\it 3})
Scalar Masses;
({\it 4}) Gaugino Masses; ({\it 5}) Trilinear Scalar Couplings, starting from
the supersymmetry breaking scale down to some intermediate scale $M_I$ at
which some scalar mass turns negative, and the gauge symmetry $[{\rm SU(3)}]^3$
can be broken down to the standard model gauge group by allowing particles with
$[{\rm SU(3)}]^3$ quantum numbers (1,3,2) and (1,3,3), {\it i.e.}, $\nu^c$
(${\bar \nu}^c$) and $N$ (${\bar N}$) in lepton (mirror-lepton) nonets, to grow
large vacuum expectation values (VEV).
     After intermediate symmetry breaking, some zero modes will acquire
superheavy masses, due to the large VEV growth along $\nu^c$
(${\bar \nu}^c$) and $N$ (${\bar N}$) directions, and hence will decouple
from low energy spectrum. Therefore, it is possible to work out the
low mass particles that will survive down to the electroweak scale and further
running of the renormalization group equations down to the
electroweak scale will give some
definite predictions for the low energy phenomenology.

    The renormalization group equations for SUSY with most general
soft breaking terms and
arbitrary gauge groups have been worked out at one and two loop levels by many
authors [26, 27]. It is straightforward to calculate them for
the model at hand.
For example, the renormalization group equation for ${\lambda^1}_{123}$
at one loop level is:
$$\eqalign{{d\over dt}{\lambda^1}_{123}
 &={{\lambda^1}_{123}\over16\pi^2}{\{}
   12({\lambda^1}_{123})^2+32({\lambda^1}_{122})^2+32({\lambda^1}_{133})^2
  +3({\lambda^4}_{212})^2\cr
 &+6({\lambda^4}_{213})^2+3({\lambda^4}_{313})^2
  +3({\lambda^4}_{222})^2+6({\lambda^4}_{223})^2+3({\lambda^4}_{323})^2\cr
 &+3({\lambda^4}_{232})^2+6({\lambda^4}_{233})^2+3({\lambda^4}_{333})^2
  +6({\lambda^4}_{142})^2+6({\lambda^4}_{143})^2\cr
 &+6({\lambda^4}_{241})^2+6({\lambda^4}_{341})^2
  +6({\lambda^4}_{162})^2+6({\lambda^4}_{163})^2\cr
 &+6({\lambda^4}_{172})^2+6({\lambda^4}_{173})^2
  -8({g_C}^2+{g_L}^2){\}},\cr} \eqno(16)$$
where $t=ln{\mu\over M_c}$ and the notations for Yukawa couplings follow those
of Eq. (14).

Some explanations about the coefficients in Eq.~(16) are in order.
Because supersymmetry is softly broken, the RG equations are dictated
only by the possible wave function renormalizations which are governed by the
renormalizable interactions.
There are seven different types of Yukawa couplings as indicated by
Eq.~(14): ${\lambda^1}_{ijk}$, ${\lambda^2}_{ijk}$,
${\lambda^3}_{ijk}$, ${\lambda^4}_{ijk}$,
${\bar\lambda}_{{\bar i}{\bar j}{\bar k}}$, ${\eta}_{ij{\bar k}}$,
and ${\rho}_{ijk}$.
Thus, the wave function renormalization for, say, ${Q_2}^a_l$, will receive a
contribution from, say, ${\lambda^1}_{123}$, with a coefficient determined by
$$({\lambda^1}_{123})^2\epsilon_{aa_2a_3}\epsilon_{ll_2l_3}
  \delta_{a_2a'_2}\delta_{a_3a'_3}\delta_{l_2l'_2}\delta_{l_3l'_3}
  \epsilon_{a'a'_2a'_3}\epsilon_{l'l'_2l'_3}
 =4({\lambda^1}_{123})^2\delta_{aa'}\delta_{ll'}.\eqno(17)$$
Similarly, the contribution from, say, ${\lambda^4}_{213}$ is
$$({\lambda^4}_{213})^2(\delta_{aa_3}\delta_{a_3a'_3}\delta_{a'_3a'})
(\delta_{r_1r_3}\delta_{r_1r'_1}\delta_{r_3r'_3}\delta_{r'_1r'_3})
(\delta_{ll_1}\delta_{l_1l'_1}\delta_{l'_1l'})
=3({\lambda^4}_{213})^2\delta_{aa'}\delta_{ll'}.\eqno(18)$$
The coefficient $3\times4=12$ in front of $({\lambda^1}_{123})^2$ in
Eq.~(16) arises because ${\lambda^1}_{123}$ receives
contributions from the wave function renormalization of three fields,
$Q_1$, $Q_2$ and $Q_3$.
Analogously, ${\eta}_{ij{\bar k}}$ will contribute a factor of 9 to the wave
function renormalization of $\phi_i$ and a factor of 1 to that of $L_j$
or ${\bar{L}}_k$, respectively.

In our analysis of the intermediate scale, we have used the
renormalization group equations at the one loop level, except for the gauge
couplings where we have used RG equations at the two loop level (excluding the
dependence on Yukawa couplings).
The reason the more accurate treatment of the gauge couplings is needed
arises from the fact that the ${\rm SU(3)}_C$ coupling constant is not
renormalized at the one loop level, {\it i.e.}
$${d\over dt}g_I
  =\{{1\over16\pi^2}a_I
  +{1\over(16\pi^2)^2}\sum_J b_{IJ}{g_J}^2\}{g_I}^3,   \eqno(19)$$
where
$$ a_I=\pmatrix{0\cr 18\cr 18};\qquad
b_{IJ}=\pmatrix{48&24&24\cr
                24&252&120\cr
                24&120&252\cr},
   \eqno(20)$$
and $I, J=C, L, R$ are the indices for
the color, left and right
$SU(3)$ subgroups. One also notices that $g_L$ and
$g_R$ have the same values as the results of running the renormalization group
equations, thus the
Weinberg angle is $\sin^2\theta_w=3/8$ before the intermediate breaking,
in accordance with the standard SU(5) grand unified theory.

    One must be careful not to ignore tiny mixings among generations
while running the renormalization group equations. Although the
non-renormalization theorem
guarantees that, aside from possible wave function renormalizations, there is
no need for coupling constant or mass renormalizations, the potential mixings
among generations can have tremendous effects on the running of the RG
equations, and even the determination of the direction of VEV growth.
In the presence of mixings, a scalar mass squared turning negative does not
necessarily mean that we have symmetry breaking.
One must diagonalize the scalar
mass squared matrix in generation space and find the eigenvectors with
negative eigenvalues of the scalar mass squared matrix. Only along these
directions can scalars trigger the intermediate breaking by growing VEV's.
As it turns out, {\it the mixing creates many new Yukawa couplings which would
otherwise remain zero, and these new couplings are, in general, of order
of $10^{-2}$ and even $10^{-4}$ compared to the usual ones
which are $\rm O(1)$} (Table 1).
{\it The generation mixing may provide a natural avenue to solve
the lepton/quark mass hierachy problem}\footnote{$^\star$}{This idea
            is supported by a recently
            revived interest on possible {\it ansatz} for fermion mass
            matrices in SUSY GUTs [28].
            The two parameters $\delta_u$ and $\delta_d$ in Eq.~(8) of
            Ref.~[28] arise actually from the generation mixing.}.
Notice that, for most string theories in four dimensions, the presence of the
generation mixing is a generic phenomenon\footnote{$^{\star\star}$}{The
            treatment of the renormalization group equations for
            the Tian-Yau model in Ref.~[29] ignored all the mixings.
            We have checked their results in the ${\overline{27}}$ sector
            and obtained the same results, but the breaking  takes place
            along different directions when the mixings are present.}.

    In the Gepner three generation model, the structure of the Yukawa couplings
reveals
that there are many mixings occurring. For example, ${\lambda^3}_{146}$ and
${\lambda^3}_{156}$ give rise to a mixing between $L_4$ and $L_5$
(Figure 1(a)), and ${\lambda^3}_{246}$ and ${\lambda^3}_{267}$ produce
a mixing between $L_4$ and $L_7$ (Figure 1(b)). A new coupling
${\lambda^3}_{347}$ develops from this latter mixing (Figure 2).
The renormalization group equation for ${\lambda^3}_{347}$ is
$${d\over dt}{\lambda^3}_{347}
 ={1\over16\pi^2}
   (8{\lambda^3}_{246}{\lambda^3}_{267}{\lambda^3}_{377}+\cdots),
  \eqno(21)$$
where $\cdots$ represents the many terms that are proportional to the
many new couplings listed in Table 1 which have vanishing tree level values.
One sees clearly that, even if one starts running this equation from the zero
value for ${\lambda^3}_{347}$, the nonzero values will develop from the first
term on the right-hand side.
In Table 1, we also list all the new $(27)^3$couplings arising from mixing.

The scalar mass squared matrix takes a block diagonal symmetric form
in generation space:
$$(M^2)_{ab}=(3\times3)\otimes(4\times4)\otimes(2\times2)
\otimes(2\times2),
\eqno(22)$$
where the $(3\times3)$ matrix denotes the mixing among $L_1$, $L_2$, $L_3$;
the $(4\times4)$ matrix represents that among $L_4$, $L_5$, $L_6$ and $L_7$;
and the $(2\times2)$ matrices are those among $Q_2$, $Q_3$ and among
${Q^c}_2$, ${Q_3}^c$ which have the same form and values.
In addition, these masses obey
$$\eqalign{(m^2)_{l_8}&=(m^2)_{l_9};\qquad (m^2)_{l_4}=(m^2)_{l_5};\cr
(m^2)_{l_4l_i}&=(m^2)_{l_5l_i},\qquad for~~~~i=6,7;\cr
(M^2)_{q_iq_j}&=(M^2)_{q^c_iq^c_j},\qquad for~~~~i,j=1,2,3.\cr}
\eqno(23)$$
Thus there are 14 independent mass parameters for leptons, 6 for mirror
leptons and 4 for quarks and anti-quarks.
Because there are far too many gauge singlets, we excluded all the gauge
singlets but $\phi_{45}$ and $\phi_{58}$ into the running of the RG
equations, just to demonstrate their effects on RG equations.
In the actual running, we also excluded the gaugino masses and the trilinear
scalar couplings for simplicity.

    Let's briefly summarize our results for the running of the
renormalization group equations.
Our analysis was performed in two steps: first, the gauge singlets were
ignored.
The system then decouples into two sectors: the 27 and ${\overline{27}}$
sectors.
We found that the symmetry breaking occurs at very high scales:
$$M_I\approx4.6\times10^{16}~~({\rm GeV})\eqno(24)$$
for the 27 sector, and
$$M_I\approx1.7\times10^{16}~~({\rm GeV})\eqno(25)$$
for the ${\overline{27}}$ sector. But the 27 sector breaking is along
the wrong
direction --- the lower eigenvalue of the scalar mass squared between $Q_2$ and
$Q_3$ (thus ${Q^c}_2$ and ${Q^c}_3$ also) turns negative before everything
else\footnote{$^\star$}{This is in disagreement with the results obtained
            in Ref.~[20] where a breaking along $L_3$ direction was found.
            However, this analysis ignored the contributions from the
            generation mixing we discussed before.}.
Notice that the intermediate scale for 27 sector is about three times
higher than that of ${\overline{27}}$ sector. Hence the VEV growth at
intermediate scale will automatically be {\it along quark or anti-quark
sector}, and this means that {\it $SU(3)_C$ breaks} --- a phenomenological
disaster.

    A scan of the singlet $[1(27{\overline{27}})]$ interactions reveals that
many gauge singlets couple to chiral fermions more strongly than the chiral
fermions couple among themselves. Thus it is necessary to consider these
couplings in the analysis. Another remarkable fact is that, due to the
chosen embedding of the discrete group $Z_3\times Z'_3$ into $E_6$, no mirror
generations of quarks and anti-quarks come into play. It is therefore not
totally unreasonable for gauge singlets to shift the symmetry breaking
direction.  For example, the gauge singlets $\phi_{45}$, $\phi_{58}$,
$\phi_{57}$, $\phi_{60}$ and $\phi_{61}$ have the following interactions with
leptons and mirror leptons:
$$\eqalign{
\eta_{{\underline {58}}7{\bar4}}&=15.018\quad
\eta_{{\underline {45}}2{\bar4}}=6.087,;\cr
\eta_{{\underline {57}}6{\bar1}}=5.230,&\quad
\eta_{{\underline {60}}4{\bar6}}=1.424,\quad
\eta_{{\underline {61}}5{\bar5}}=1.424.\cr
}\eqno(26)$$
in the notations of Eq. (14).
Hence, if one introduces two gauge singlets $\phi_{45}$ and $\phi_{58}$ into
the running of the renormalization group equations,
one can reasonably expect that the
mass squared of the scaler component of ${\bar L}_4$ will turn negative first,
and due to the absence of the mixing in the ${\overline{27}}$ sector and the
C-even property of ${\bar L}_4$, ${\bar N}_4$ alows VEV growth
and thus triggers the symmetry breaking.

There are still some subtleties arising due to the unknown origins of
supersymmetry breaking. As mentioned earlier, we assume some mechanism to
break the supersymmetry which in turn introduces a common mass for all the
scalar particles. But because of the uncertainty involved in SUSY breaking, we
may assign different masses to scalar particles belonging to different
representations of the gauge group, i.e., the gauge singlets may acquire masses
different from those of 27 or ${\overline{27}}$ scalars. Therefore, in running
the renormalization group equations, we kept all the gauge
nonsinglet masses fixed, and gauge
singlet masses was entered as a free parameter.
Defining $R={m^2}_{singlet}/{m^2}_{non-singlet}$, we considered the range
$0~<~R\leq100.$ For $R\sim O(1)$, the gauge singlet mass turns negative rather
rapidly, at a scale very close to the compactification scale
$${\rm M_{singlet}}\sim2.35\times10^{18}~\rm GeV, \eqno(27)$$
because of the large values of $[1(27{\overline{27}})]$ couplings.
For $R\geq7$, the ${\bar L}_4$ mass squared turns negative at a scale of
$${\rm M'_{singlet}}\sim2.19\times10^{18}~\gev,\eqno(28)$$
For $R>50$, the ${\bar L}_4$
mass squared turns negative at a scale larger than that
of the case when $R\sim O(1)$.
Of course, one may suspect the validity of the renormalization group
equation approach since the
scalar masses turn negative almost immediately due to the large value of
$[1(27{\overline{27}})]$ couplings.
We think that some non-perturbative techniques must be developed before
one can properly deal with this kind of problems.

There exists the possibility that these singlets grow VEV's first
and give superheavy masses to some particles, and this may prevent us from
running into the afore-mentioned problem of having the color \su3 subgroup
broken.
In order to explore this possibility,
    we assume a four step scenario in which the gauge singlets
grow VEV's in the following order:
\item{}({\it i}) $\phi_{58}$ grows VEV, then $L_7$ and ${\bar L}_4$ pair up
and decouple;
\item{}({\it ii}) $\phi_{57}$ grows VEV, then $L_6$ and ${\bar L}_1$ pair up
and decouple;
\item{}({\it iii}) $\phi_{60}$ grows VEV, then $L_4$ and ${\bar L}_6$ pair up
and decouple;
\item{}({\it iv}) $\phi_{61}$ grows VEV, then $L_5$ and ${\bar L}_5$ pair up
and decouple.

\noindent Note that the last two steps may be interchanged. Our choice of
having $\phi_{58}$ grow VEV rather than $\phi_{45}$ in ({\it i}) is because
the survey on this model suggested that the possible light Higgs, after
including the loop corrections, should necessarily lie in a direction along
which the first three lepton generations mix [30]. Hence we do expect
$L_4$, $L_5$, $L_6$ and $L_7$, but not $L_2$,
to become superheavy and decouple. We also notice that in the scenario
suggested above, four pairs of leptons and mirror leptons decouple. The
four decoupled pairs are all C-even states (Sec. 4),  except ${\bar L}_1$
which is C-odd. In view of the results we have already obtained, we can
reasonably expect ({\it i}) -- ({\it iv}) to happen at a very large scale
$M_{singlet}\approx5.0\times10^{17}$ GeV. After the decoupling of four
generations of lepton-mirror lepton pairs, we are left with 5 lepton
($L_i,~i=1,2,3,8,9$), 2 mirror lepton (${\bar L}_{\bar i},~{\bar i}=2,3$),
and 3 intact quark and conjugate quark generations. Of all 52 independent
Yukawa couplings, there are only 15 survived (Table 1).
The values of these Yukawa couplings in the second column in Table 1 are
the values obtained from the running of the renormalization group equations
when only two gauge singlets are involved, which
give a rough picture about their sizes. We can run the
renormalization group equations with these
nonets and Yukawa couplings and it can be expected that the gauge symmetry
breaking will
take place at a scale, say, greater than $1.0\times10^{16}$ GeV.
The low-energy spectrum and the proton stability will be sudied in a
subsequent paper.

\centerline{\bf 4.~~Conclusions}

    We have analysed here the Gepner three generation heterotic string model,
examining {\it via} the
renormalization group equations the intermediate scale symmetry breaking.
As expected, the model does demonstrate a very large intermediate breaking
scale of O($1.0\times10^{16}$ GeV) or larger. Actually, a very large
intermediate scale may be expected to be a generic feature of the four
dimensional effective field theory of heterotic string theory. This is because
this class of models in general have a very large number of renormalizable
Yukawa couplings which can rapidly turn a $(mass)^2$ negative.

 There are a few lessons that we have learned which we briefly discuss:
    In analysing the
renormalization group equations of a coupled system as complex as the one we
have studied and the one studied in [29], one must be very careful not to
discard the presumably small mixing terms like those shown in Fig.(1).
In fact, from our study of the Gepner three generation
model, we conclude that these mixing terms in generation space play a very
important role in extracting the phenomenological implications of the model.
They can trigger the direction of scalar field VEV growth. They also offer
the possibility of solving the generation mass hierarchy problem by producing
additional very small Yukawa couplings among chiral families, which are zero
at tree level but grow due to  loop corrections. The origin of these is the
presence of mixing among generations in the wavefunction renormalizations. We
have also seen the importance of the gauge singlet couplings. We found that
they provide almost a unique source to prevent the very early breaking of color
gauge group in the Gepner three generation model.

\vskip 10truept
    We thank Stefan Cordes for sending us reference [22] prior to the
publication, and one of us (J. W.) thanks him for helpful discussions.
This work is supported in part under National Science Foundation Grant No.
PHYS-916593.
\vskip 10truept
\centerline{\bf References}

\item{[1]}  M. B. Green and J. H. Schwarz, \PLB{149}{84}{117}.
\item{[2]}  D. J. Gross, J. A. Harvey, E. Martinec and R. Rohm,
            \PRL{54}{85}{502}; \NPB{256}{85}{253}.
\item{[3]}  G. Tian and S. T. Yau, {\it in} Proc. of Argonne Symposium,
            ``Anomalies, Geometry and Topology'' eds. W. A. Bardeen and
            A. White (World Scientific, Singapore, 1985).
\item{[4]}  P. Candelas, G. Horowitz, A. Strominger and E. Witten,
            \NPB{258}{85}{47}.
\item{[5]}  D. Gepner, \NPB{296}{88}{757}. For a review, see
            D. Gepner, Lectures on $N=2$ string theory, {\it in}
            SUPERSTRINGS'89, Proceedings of the Trieste Spring School,
            April, 1989. eds. M. Green, R. Iengo, S. Randjbar-Daemi, E. Sezgin
            and A. Strominger (World Scientific, 1990).
\item{[6]}  E. Martinec, \PLB{217}{88}{431}, and
            {\it in} Physics and Mathematics of Strings,
            Memorial Volume for Vadim Knizhnik, eds. L. Brink, D. Friedan and
            A. M. Polyakov (World Scientific, 1990);
\item{}     W. Lerche, C. Vafa and N. P. Warner, \NPB{324}{89}{427};
\item{}     B. R. Greene, C. Vafa and N. P. Warner, \NPB{328}{89}{371}.
\item{[7]}  B. R. Greene and M. R. Plesser, \NPB{338}{90}{15};
\item{}     P. S. Aspinwall, C. A. L\"utken and G. G. Ross, \PLB{241}{90}{373}.
\item{}     P. Candelas, X. C. de la Ossa, P. S. Green and L. Parks,
            \NPB{359}{91}{21}, \PLB{258}{91}{118}.
\item{[8]}  B. R. Greene, K. H. Kirklin, P. J. Miron and G. G. Ross,
            \PLB{180}{86}{69}, \NPB{278}{86}{667},
            {\it ibid.} {B292} 606 (1987);
\item{}     P. Nath and R. Arnowitt,  \PRL{62}{89}{1437}, \PRD{39}{89}{2006};
\item{}     R. Arnowitt and P. Nath,  \PRL{62}{89}{2225}, \PRD{40}{89}{191}.
\item{[9]}  R. Schimmrigk,  \PLB{193}{87}{175}.
\item{[10]} D. Gepner,  Princeton University Preprint, (December, 1987).
\item{[11]} P. Candelas, \NPB{298}{88}{458}.
\item{[12]} M. Dine, N. Seiberg, X. G. Wen and E. Witten, \NPB{278}{86}{769},
            {\it ibid.}, {B289} (1987) 319;
\item{}     M. Dine and N. Seiberg, \NPB{306}{88}{137}.
\item{[13]} J. Distler and B. R. Greene, \NPB{309}{88}{295};
\item{}     D. Gepner, \NPB{311}{88}{191}.
\item{[14]} P. DiVechia, J. L. Peterson and H. B. Zheng, \PLB{174}{86}{280};
\item{}     W. Boucher, D. Friedan and A. Kent,  \PLB{172}{86}{316};
\item{}     Z. Qui, \PLB{188}{87}{207}.
\item{[15]} A. B. Zamolodchikov and V. A. Fateev, Sov. Phys. JETP {62}
            (1985) 215;
\item{}     D. Gepner and Z. Qui, \NPB{285[FS19]}{87}{423}.
\item{[16]} J. Wess and B. Zumino, \PLB{37}{71}{95};
\item{}     E. Witten,  Commun. Math. Phys. {92} (1984) 445.
\item{[17]} V. G. Knizhnik and A. B. Zamolodchikov,  \NPB{247}{84}{83};
\item{}     A. B. Zamolodchikov and V. A. Fateev,  Sov. J. Nucl. Phys. {43}
            (1986) 657.
\item{[18]} E. Witten, \NPB{258}{85}{79}.
\item{[19]} J. Wu, R. Arnowitt and P. Nath, \IJMPA{6}{91}{381}.
\item{[20]} G. G. Ross, Proceedings of {\it Strings'89} workshop, eds.
\item{}     R. Arnowitt, R. Bryan, M. Duff, D. V. Nanopolous and C. N. Pope
            (World Scientific, Singapore, 1990.)
\item{[21]} B. R. Greene, R. M. Plesser, E. Rusjan and X. M. Wang,
            \MPLA{6}{91}{591}.
\item{[22]} S. Cordes and Y. Kikuchi, \IJMPA{6}{91}{5017};
\item{}     S. Cordes,  private communications.
\item{[23]} S. Cordes and Y. Kikuchi,  Texas A\&M Preprints,
            CTP-TAMU-92/88 (December, 1988);
            and \MPLA{4}{89}{1365}.
\item{[24]} A. Kato and Y. Kitazawa, \NPB{319}{89}{474};
\item{}     M. R. Douglas and S. P. Trevidi, \NPB{320}{89}{461};
\item{}     J. Fuchs and A. Klemm, Ann. Phys. {194} (1989) 303;
\item{}     B. R. Greene, C. A. L\"utken and G. G. Ross, \NPB{325}{89}{101}.
\item{[25]} M. Dine, V. Kaplunovsky, M. Mangano, C. Nappi and N. Seiberg,
            \NPB{259}{85}{549};
\item{}     J. Ellis, K. Enquist, D. V. Nanopolous and K. Olive,
            \PLB{188}{87}{415};
\item{}     R. Arnowitt and P. Nath, \PRL{60}{88}{1817}.
\item{[26]} K. Inoue, A. Kakuto, H. Komatsu and S. Takeshita, Prog. Theor.
            Phys. {68} (1982) 927, {\it ibid.} {71} (1984) 413;
\item{}     L. Alvarez-Gaum\'e, J. Polchinski and M. B. Wise,
            \NPB{221}{83}{495};
\item{}     N. K. Falck, Z. Phys. {C30} (1986) 247.
\item{[27]} D. R. T. Jones, \PRD{25}{581}{82};
\item{}     J. E. Bj\"orkman and D. R. T. Jones, \NPB{259}{85}{533}.
\item{[28]} S. Dimopoulos, L. J. Hall and S. Raby, \PRD{45}{92}{4192}.
\item{[29]} F. del Aguila and G. D. Coughlan, \PLB{215}{88}{93}.
\item{[30]} R. Arnowitt and P. Nath, {\it in} Proceedings of {\it Strings'90}
            workshop,
            eds. R. Arnowitt, R. Bryan, M. Duff, D. V. Nanopolous, C. N. Pope
            and E. Sezgin (World Scientific, Singapore, 1991.)
\vskip 10truept
\centerline{\bf Tables and Figures}
\centerline{Table~~1.~~~Yukawa~Couplings}

This table contains all the Yukawa couplings we considered in the text when
analyzing the renormalization group equations. The first and fourth columns are
the couplings whose notations follow those of Eq. (14), the second and fifth
columns are the tree level values of the corresponding couplings and the
third and sixth are their one-loop level values. In obtaining these values,
we have taken the following initial conditions: the common gauge coupling
constant, $g_0=0.70$, the compactificatin scale,
${\rm M_c}=2.4\times10^{18}$ \gev, and the gauge singlet and the gauge
non-singlet mass ratio, $R=10.0$. Then, it was found that
${\rm M_I}\sim1.91\times10^{18}$ \gev and the ${\bar L}_4$
mass turns negative first.

\cl{\bf Figure Captions}
\vskip 10pt
\ni{Fig. 1(a)} The mixing between $L_4$ and $L_5$.

\ni{Fig. 1(b)} The mixing between $L_4$ and $L_7$.

\ni{Fig. 2~~~~} The new coupling ${\lambda^3}_{347}$ arises from the mixing
             in Fig. 1(b).
\vfill
\eject
\nopagenumbers
\vbox{\offinterlineskip
      \hrule
      \halign{&\vrule#&
      \strut\hfil#\hfil\hfil\hfil\hfil\cr
     height2pt&\omit&&\omit&&\omit&&\omit&&\omit&&\omit&\cr
     &~~Coupling~~\hfil&&~~Tree Value~~&&~~One-loop Value~~&
     &~~Coupling~~&&~~Tree Value~~&&~~One-loop Value~~&\cr
     \noalign{\hrule}
    height2pt&\omit&&\omit&&\omit&&\omit&&\omit&&\omit&\cr
      &${\lambda^1}_{122}$&&-&&$-2.470\times10^{-3}$&
      &${\lambda^2}_{122}$&&-&&$-2.470\times10^{-3}$&\cr
    height2pt&\omit&&\omit&&\omit&&\omit&&\omit&&\omit&\cr
      &${\lambda^1}_{123}$&&0.654&&0.629&
      &${\lambda^2}_{123}$&&0.654&&0.629&\cr
    height2pt&\omit&&\omit&&\omit&&\omit&&\omit&&\omit&\cr
      &${\lambda^1}_{133}$&&0.537&&0.511&
      &${\lambda^2}_{133}$&&0.537&&0.511&\cr
     \noalign{\hrule}
    height2pt&\omit&&\omit&&\omit&&\omit&&\omit&&\omit&\cr
      &${\lambda^3}_{144}$&&-&&$-8.133\times10^{-4}$&
      &${\lambda^3}_{155}$&&-&&$-8.133\times10^{-4}$&\cr
    height2pt&\omit&&\omit&&\omit&&\omit&&\omit&&\omit&\cr
      &${\lambda^3}_{146}$&&0.577&&0.559&
      &${\lambda^3}_{156}$&&0.577&&0.559&\cr
    height2pt&\omit&&\omit&&\omit&&\omit&&\omit&&\omit&\cr
      &${\lambda^3}_{147}$&&-&&$~~1.330\times10^{-5}$&
      &${\lambda^3}_{157}$&&-&&$~~1.330\times10^{-5}$&\cr
    height2pt&\omit&&\omit&&\omit&&\omit&&\omit&&\omit&\cr
      &${\lambda^3}_{145}$&&-&&$-1.603\times10^{-3}$&
      &${\lambda^3}_{166}$&&-&&$-1.625\times10^{-3}$&\cr
    height2pt&\omit&&\omit&&\omit&&\omit&&\omit&&\omit&\cr
      &${\lambda^3}_{167}$&&-&&$-3.829\times10^{-3}$&
      &${\lambda^3}_{177}$&&-&&$~~1.543\times10^{-5}$&\cr
    height2pt&\omit&&\omit&&\omit&&\omit&&\omit&&\omit&\cr
      &${\lambda^3}_{189}$&&-&&$-1.868\times10^{-3}$&
      &${\lambda^3}_{245}$&&-&&$-2.605\times10^{-3}$&\cr
    height2pt&\omit&&\omit&&\omit&&\omit&&\omit&&\omit&\cr
      &${\lambda^3}_{244}$&&-&&$-6.438\times10^{-4}$&
      &${\lambda^3}_{255}$&&-&&$-6.438\times10^{-4}$&\cr
    height2pt&\omit&&\omit&&\omit&&\omit&&\omit&&\omit&\cr
      &${\lambda^3}_{246}$&&0.475&&0.440&
      &${\lambda^3}_{256}$&&0.475&&0.440&\cr
    height2pt&\omit&&\omit&&\omit&&\omit&&\omit&&\omit&\cr
      &${\lambda^3}_{247}$&&-&&$-9.075\times10^{-4}$&
      &${\lambda^3}_{257}$&&-&&$-9.075\times10^{-4}$&\cr
    height2pt&\omit&&\omit&&\omit&&\omit&&\omit&&\omit&\cr
      &${\lambda^3}_{266}$&&-&&$-1.286\times10^{-3}$&
      &${\lambda^3}_{267}$&&0.676&&0.625&\cr
    height2pt&\omit&&\omit&&\omit&&\omit&&\omit&&\omit&\cr
      &${\lambda^3}_{277}$&&-&&$-8.628\times10^{-4}$&
      &${\lambda^3}_{289}$&&0.635&&0.630&\cr
    height2pt&\omit&&\omit&&\omit&&\omit&&\omit&&\omit&\cr
      &${\lambda^3}_{344}$&&-&&$-2.460\times10^{-3}$&
      &${\lambda^3}_{355}$&&-&&$-2.460\times10^{-3}$&\cr
    height2pt&\omit&&\omit&&\omit&&\omit&&\omit&&\omit&\cr
      &${\lambda^3}_{346}$&&-&&$-1.888\times10^{-3}$&
      &${\lambda^3}_{356}$&&-&&$-1.888\times10^{-3}$&\cr
    height2pt&\omit&&\omit&&\omit&&\omit&&\omit&&\omit&\cr
      &${\lambda^3}_{347}$&&-&&$-3.198\times10^{-3}$&
      &${\lambda^3}_{357}$&&-&&$-3.198\times10^{-3}$&\cr
    height2pt&\omit&&\omit&&\omit&&\omit&&\omit&&\omit&\cr
      &${\lambda^3}_{345}$&&0.822&&0.799&
      &${\lambda^3}_{367}$&&-&&$-1.039\times10^{-3}$&\cr
    height2pt&\omit&&\omit&&\omit&&\omit&&\omit&&\omit&\cr
      &${\lambda^3}_{377}$&&0.556&&0.530&
      &${\lambda^3}_{389}$&&-&&$-1.038\times10^{-3}$&\cr
     \noalign{\hrule}
    height2pt&\omit&&\omit&&\omit&&\omit&&\omit&&\omit&\cr
      &${\lambda^4}_{212}$&&-&&$-6.040\times10^{-3}$&
      &${\lambda^4}_{313}$&&-&&$-5.408\times10^{-3}$&\cr
    height2pt&\omit&&\omit&&\omit&&\omit&&\omit&&\omit&\cr
      &${\lambda^4}_{213}$&&0.577&&0.560&
      &${\lambda^4}_{312}$&&0.577&&0.560&\cr
    height2pt&\omit&&\omit&&\omit&&\omit&&\omit&&\omit&\cr
      &${\lambda^4}_{222}$&&0.577&&0.545&
      &${\lambda^4}_{323}$&&0.390&&0.358&\cr
    height2pt&\omit&&\omit&&\omit&&\omit&&\omit&&\omit&\cr
      &${\lambda^4}_{223}$&&-&&$-5.189\times10^{-3}$&
      &${\lambda^4}_{322}$&&-&&$-5.189\times10^{-3}$&\cr
    height2pt&\omit&&\omit&&\omit&&\omit&&\omit&&\omit&\cr
      &${\lambda^4}_{232}$&&-&&$-9.077\times10^{-4}$&
      &${\lambda^4}_{333}$&&1.054&&1.005&\cr
    height2pt&\omit&&\omit&&\omit&&\omit&&\omit&&\omit&\cr
      &${\lambda^4}_{233}$&&-&&$-3.925\times10^{-3}$&
      &${\lambda^4}_{332}$&&-&&$-3.925\times10^{-3}$&\cr
    height2pt&\omit&&\omit&&\omit&&\omit&&\omit&&\omit&\cr
      &${\lambda^4}_{142}$&&0.577&&0.564&
      &${\lambda^4}_{251}$&&0.577&&0.564&\cr
    height2pt&\omit&&\omit&&\omit&&\omit&&\omit&&\omit&\cr
      &${\lambda^4}_{241}$&&-&&$-3.526\times10^{-3}$&
      &${\lambda^4}_{152}$&&-&&$-3.526\times10^{-3}$&\cr
    height2pt&\omit&&\omit&&\omit&&\omit&&\omit&&\omit&\cr
      &${\lambda^4}_{143}$&&-&&$-4.630\times10^{-3}$&
      &${\lambda^4}_{351}$&&-&&$-4.630\times10^{-3}$&\cr
    height2pt&\omit&&\omit&&\omit&&\omit&&\omit&&\omit&\cr
      &${\lambda^4}_{341}$&&0.475&&0.457&
      &${\lambda^4}_{153}$&&0.475&&0.457&\cr
    height2pt&\omit&&\omit&&\omit&&\omit&&\omit&&\omit&\cr
      &${\lambda^4}_{162}$&&-&&$-3.601\times10^{-3}$&
      &${\lambda^4}_{261}$&&-&&$-3.601\times10^{-3}$&\cr
    height2pt&\omit&&\omit&&\omit&&\omit&&\omit&&\omit&\cr
      &${\lambda^4}_{163}$&&0.740&&0.711&
      &${\lambda^4}_{361}$&&0.740&&0.711&\cr
    height2pt&\omit&&\omit&&\omit&&\omit&&\omit&&\omit&\cr
      &${\lambda^4}_{172}$&&-&&$-9.608\times10^{-4}$&
      &${\lambda^4}_{271}$&&-&&$-9.608\times10^{-4}$&\cr
    height2pt&\omit&&\omit&&\omit&&\omit&&\omit&&\omit&\cr
      &${\lambda^4}_{173}$&&-&&$-7.709\times10^{-4}$&
      &${\lambda^4}_{371}$&&-&&$-7.709\times10^{-4}$&\cr
     \noalign{\hrule}
    height2pt&\omit&&\omit&&\omit&&\omit&&\omit&&\omit&\cr
      &${\bar\lambda}_{134}$&&1.153&&1.072&
      &${\bar\lambda}_{244}$&&0.556&&0.490&\cr
    height2pt&\omit&&\omit&&\omit&&\omit&&\omit&&\omit&\cr
      &${\bar\lambda}_{256}$&&0.822&&0.812&
      &$\eta_{{\underline{45}}1{\bar4}}$&&-&&$-1.297\times10^{-2}$&\cr
    height2pt&\omit&&\omit&&\omit&&\omit&&\omit&&\omit&\cr
      &$\eta_{{\underline{45}}2{\bar4}}$&&6.081&&3.804&
      &$\eta_{{\underline{45}}3{\bar4}}$&&-&&$-7.212\times10^{-3}$&\cr
    height2pt&\omit&&\omit&&\omit&&\omit&&\omit&&\omit&\cr
      &$\eta_{{\underline{58}}4{\bar4}}$&&-&&$-4.177\times10^{-3}$&
      &$\eta_{{\underline{58}}5{\bar4}}$&&-&&$-4.177\times10^{-3}$&\cr
    height2pt&\omit&&\omit&&\omit&&\omit&&\omit&&\omit&\cr
      &$\eta_{{\underline{58}}6{\bar4}}$&&-&&$~~3.803\times10^{-5}$&
      &$\eta_{{\underline{58}}7{\bar4}}$&&2.761&&2.375&\cr
    height2pt&\omit&&\omit&&\omit&&\omit&&\omit&&\omit&\cr}\hrule}
\vfill
\eject
\end
\bye